\documentclass[onecolumn,secnumarabic,amssymb, amsmath, nofootinbib,tightenlines,
nobibnotes, aps, prl,epsfig]{revtex4}
\usepackage{graphicx}
\usepackage{dcolumn}
\usepackage{bm}
\begin{document}
\preprint{APS/123-QED}
\title{The unintegrated gluon distribution from the GBW and BGK models }

\author{G.R.Boroun}%
 \email{boroun@razi.ac.ir }
 \affiliation{ Department of Physics,
Razi University, Kermanshah
67149, Iran}
\date{\today}
\begin{abstract}
The gluon distribution is obtained from the
Golec-Biernat-W$\ddot{\mathrm{u}}$sthoff (GBW) and Bartels,
Golec-Biernat and Kowalski (BGK)  models at low $x$. We derive
analytical results for the unintegrated color dipole gluon
distribution function at small transverse momentum, which provides
useful information to constrain the $k_{t}$-shape of the
unintegrated gluon distribution in comparison with the
unintegrated gluon distribution (UGD) models. The longitudinal
proton structure function $F_{L}(x,Q^2)$ from the $k_{t}$
factorization scheme, using the unintegrated gluon density is
computed. We compare the predictions for the on-shell and twist-2
corrections with the HERA data and the CJ15 parametrization method
for $F_{L}$. We show that this method is very well described the
experimental data within the on-shell and twist-2 framework.
Effects of parameters on $F_{L}$, where charm contribution is
taken into account, are investigated. These results
are in good agreement with the data at fixed $W$.\\

\end{abstract}
 \pacs{***}
\keywords{****} 
\maketitle
\subsection{I. Introduction}

The saturation model [1] was shown some years ago and has provided
a successful description of HERA deep inelastic scattering (DIS)
data. The gluon saturation effects in the HERA data at very low
$x$ values has been discussed extensively in the context of the
color dipole model (CDM) [2]. The main measured low $x$ effect is
a strong rise of the parton distribution functions in the limit
$x{\rightarrow}0$ (for fixed virtuality $Q^2$), where the computed
cross section violates unitarity. The recombination effects at low
$x$ are responsible for saturation of the parton densities by
taming their strong rise. An important result of saturation is the
existence of a saturation scale which is reflected in a new
scaling law for inclusive DIS cross section. The saturation scale
$Q_{s}$ increases with decreasing $x$. The concept of the
collinear parton distribution functions (PDFs) is extended for
gluons, known from the collinear factorization. The unintegrated
gluon distributions (UGDs) are usually called Transverse Momentum
Dependent (TMD) gluon distributions and, indeed, their studies are
important in the context of more exclusive observables, like
correlations in the Drell-Yan pair production. While the TMD gluon
distributions have precise definitions within QCD in terms of
hadronic matrix elements of a bilocal gluon operator, the UGDs are
nowadays considered as rather vague objects [3].\\
The strong rise of the gluon distribution is predicted by the
Dokshitzer-Gribov-Lipatov-Altarelli-Parisi (DGLAP) equations [4],
which appear in the double logarithmic limit, where large
logarithms $(\alpha_{s}{\ln}(1/x){\ln}Q^2)^{n}$ have to be resumed
in the Bjorken limit ( $Q^2{\rightarrow}\infty$ at fixed $x$). In
addition to the Bjorken limit, there is  the Regge limit
($x{\rightarrow}0$ at fixed $Q^2$) where in this limit the
center-of-mass energy of the $\gamma^{*}p$ system $\sqrt{s}$ goes
to infinity since $s=Q^2/x$. An expansion in powers of $x$ (i.e.,
resummation of large logarithms $(\alpha_{s}\ln(1/x))^n$ in the
Regee limit in QCD) is to be performed and gluon saturation is
expected to emerge for very small values of $x$. The result is
given in terms of the Balitsky- Fadin-Kuraev-Lipatov (BFKL)
equation [5]
\begin{eqnarray}
\frac{{\partial}f(x,k^{2}_{t})}{{\partial}{\ln}({1}/{x})}=\int{dk'^{2}_{t}}K(k^{2}_{t},k'^{2}_{t})f(x,k'^{2}_{t}),
\end{eqnarray}
where $K$ is the BFKL kernel and $f(x,k^{2}_{t})$ is the
unintegrated gluon distribution. The BFKL leads to the UGD rising
as a power of $x$,
\begin{eqnarray}
f(x,k^{2}_{t}){\sim}h(k^{2}_{t})x^{-\lambda},
\end{eqnarray}
where $h{\sim}(k^{2}_{t})^{-\frac{1}{2}}$ at large $k^{2}_{t}$ and
$\lambda$ is the maximum eigenvalue of the kernel $K$ of the BFKL
equation.  $k_{t}$ being the transverse momentum of gluon.  For
fixed $\alpha_{s}$, $\lambda$ has the value
$\lambda=\frac{\alpha_{s}}{\pi}12\ln2$, where this hard Pomeron
has been termed the BFKL Pomeron and lead to very steeply rising
$V^{*}N$ cross-sections. In contrast to DGLAP evolution where the
dominant contribution arises from diagrams that connect the target
to the photon with a strong ordering from small to large
transverse momenta $\mu{\ll}k_{t,1}{\ll}k_{t,2}{\ll}...{\ll}Q$,
the transverse components are assumed to be of the same order,
i.e., $k_{t,1}{\sim}k_{t,2}{\sim}...{\sim}k_{t,n}$ along the
cascade [6] in the BFKL kinematics. The function $f(x,k^{2}_{t})$
is related to the gluon distribution in the double logarithmic
limit by the following form
\begin{eqnarray}
xg(x,Q^{2})&{\equiv}&\int^{Q^{2}}\frac{dk_{t}^{2}}{k_{t}^{2}}f(x,k_{t}^{2}).
\end{eqnarray}
Using the $k_{t}$-factorization theorem which contains all
twists\footnote{In the twist expansion, the large logarithms
$\ln(1/x)$ at low $x$ are important as
\begin{eqnarray}
F_{2}(x,Q^2)=F^{(0)}_{2}(x,\ln{Q^2})+F^{(1)}_{2}(x,\ln{Q^2})\frac{M^2}{Q^2}...,\nonumber
\end{eqnarray}
where the $k_{t}$-factorization theorem is necessary allowing
these large logarithms independent of the twist expansion [3].},
the structure function at low $x$ is determined by
\begin{eqnarray}
F_{2}(x,Q^2)=\int\frac{dk_{t}^{2}}{k^{2}_{t}}\Phi(Q^2,k_{t})f(x,k_{t}),
\end{eqnarray}
where $\Phi(x,k_{t})$ is the virtual photon impact factor
describing the process
$\gamma^{*}{\rightarrow}q\overline{q}{\rightarrow}\gamma^{*}$.
Using the Fourier conjugate of the $k_{t}$-factorization formula
(4), the $F_{2}$ structure function
\begin{eqnarray}
F_{2}(x,Q^2)=\frac{Q^2}{4\pi^2\alpha_{em}}\bigg{(}\sigma^{\gamma^{*}p}_{T}(x,Q^2)+\sigma^{\gamma^{*}p}_{L}(x,Q^2)\bigg{)},
\end{eqnarray}
where the transverse size of gluons with transverse momentum
$\mathbf{k}$ is proportional to $1/|\mathbf{k}|$. Indeed the
transverse momentum $\mathbf{k}$ is traded for its conjugate
transverse separation $\mathbf{r}$. The measured longitudinal
structure function is related to $\sigma_{L}^{\gamma^{*}p}(x,Q^2)$
by the standard formula
\begin{eqnarray}
F_{L}(x,Q^2)=\frac{Q^2}{4\pi^2\alpha_{em}}\sigma^{\gamma^{*}p}_{L}(x,Q^2).
\end{eqnarray}
It is well known that the scattering between the virtual photon
$\gamma^{*}$ and the proton is seen as the dissociation of
$\gamma^{*}$ into a $q\overline{q}$ pair (the color dipole)
following by the interaction of this dipole with the color fields
in the proton [7], as the $\gamma^{*}p$ cross-sections are defined
by the following forms
\begin{eqnarray}
\sigma_{L,T}^{\gamma^{*}p}(x,Q^{2})=\int dz d^{2}\mathbf{r}
|\Psi_{L,T}(\mathbf{r},z,Q^{2})|^{2}\sigma_{\mathrm{dip}}({x},\mathbf{r}).
\end{eqnarray}
 Here
 $\sigma_{\mathrm{dip}}({x},r)$ is the
dipole cross-section which related to the imaginary part of the
$(q\overline{q})p$ forward scattering amplitude as the transverse
dipole size $r$ and the
 longitudinal momentum fraction $z$ with respect to the photon
 momentum are defined. The variable $z$, with $0\leq z \leq 1 $, characterizes the
distribution of the momenta between quark and antiquark. In
Eq.(7), $\Psi_{L,T}$ are the appropriate spin averaged light-cone
wave functions of the photon as the square of the photon wave
function describes the probability for the occurrence of a
$(q\overline{q})$ fluctuation of transverse size with respect to
the photon polarization. The light-cone photon wave function,
$\Psi$ is modelled by the lowest order
$\gamma^{*}g{\rightarrow}q\overline{q}$ scattering amplitudes
which give
\begin{eqnarray}
|\Psi_{T}^{f}(z,r,Q^2)|^2=\frac{2N_{c}\alpha_{em}e_{f}^{2}}{4\pi^2}\bigg{\{}
[z^2+(1-z)^2]\epsilon^2K_{1}^{2}(\epsilon{r})+m_{f}^{2}K_{0}^{2}(\epsilon{r})\bigg{\}}
\end{eqnarray}
and
\begin{eqnarray}
|\Psi_{L}^{f}(z,r,Q^2)|^2=\frac{8N_{c}\alpha_{em}e_{f}^{2}}{4\pi^2}
Q^2z^2(1-z)^2K_{0}^{2}(\epsilon{r}),
\end{eqnarray}
where $\epsilon^2=z(1-z)Q^2+m_{f}^{2}$, $K_{0}$ and $K_{1}$ are
modified Bessel functions and the sum is over quark flavors $f$
with quark mass $m_{f}$.\\
Although our knowledge of the proton structure at low $x$ is very
limited. However the data from HERA has enabled us to answer many
questions in that domain and largely improved our picture of
interior of a proton. The color dipole picture (CDP) has been
introduced to study a wide variety of low $x$ inclusive and
diffractive processes at HERA and gives a clear interpretation of
the high-energy interactions. The CDP is characterized by high
gluon densities because the proton structure is dominated by dense
gluon systems [8-10] and predicts that the low $x$ gluons in a
hadron wave function should form a Color Glass Condensate [11].
The next generation of DIS machines (i.e., the Election Ion
Collider (EIC) [12] and the Large Hadron Electron Collider (LHeC)
[13])  are making its way and will soon allow us to uncover more
details about hadron structure.\\
In the present study, we will use
the model proposed by Golec-Biernat and W$\mathrm{\ddot{u}}$sthoff
(GBW) [1] and its extension by Bartels, Golec-Biernat and Kowalski
(BGK) [8]. These models allows us to relatively easily investigate
the role of the exact gluon density. The resulting of the GBW+BGK
model is used in DIS at LHeC. This method will compare to the
results in Ref.[14], where the gluon distribution is obtained from
the dipole fits to include the contribution from heavy flavors to
the inclusive structure function. The author in Ref.[14] has
investigated the relationship between the gluon distribution
obtained using a dipole model fit to small $x$ data on
$F_{2}(x,Q^2)$ and standard gluons obtained from global fits with
the collinear factorization theorem at fixed
order.\\
The paper is organized as follows. In the next section, we present
theoretical framework for the saturation models in the
$k_{t}$-factorization and the dipole factorization. In Sec. III,
results of our study in the $k_{t}$-factorization formula are
presented. We apply the results from the UGD to compute the
longitudinal structure function  in DIS  and make comparisons to
the HERA data. Section IV contains conclusions.\\

\subsection{II. Saturation Models}

The function $\sigma_{\mathrm{dip}}(x,\mathbf{r})$ in Eq.(7) is
the color dipole cross section and give a good description of the
inclusive total $\gamma^{*}p$ cross section. In the GBW model [1],
the dipole cross-section depends on the dipole size $r$ and the
Bjorken variable $x$, and takes the following form
\begin{eqnarray}
\sigma_{\mathrm{dip}}(x,\mathbf{r})=\sigma_{0}\bigg{\{}1-
\exp\bigg{(}r^2Q_{\mathrm{sat}}^2(x)/4\bigg{)} \bigg{\}},
\end{eqnarray}
where $Q_{\mathrm{sat}}(x)$ plays the role of the
saturation\footnote{Saturation is visible in the fact that the
dipole scattering amplitude approaches the unitarity bound for the
dipole sizes larger than a characteristic size $1/Q_{s}(x)$ which
decreases when decreasing $x$ [7]. } momentum, parametrized as
$Q_{\mathrm{sat}}^2(x)=(x_{0}/x)^\lambda \mathrm{GeV}^2 $. The
parameters of the model (i.e.,$\sigma_{0}$, $x_{0}$ and $\lambda$
) are found from a fit to small-$x$ data [15]. At small $r$,
$\sigma_{\mathrm{dip}}$ features colour transparency,
$\sigma_{\mathrm{dip}}{\sim}r^2$, which is purely pQCD phenomenon,
while for large $r$, $\sigma_{\mathrm{dip}}$ saturates,
$\sigma_{\mathrm{dip}}{\simeq}\sigma_{0}$ [3]. Since the photon
wave function depends on mass of the quarks in the $q\overline{q}$
dipole, the Bjorken variable $x$ is modified \footnote{This is the
formal photoproduction limit.} by the following form [15]
\begin{eqnarray}
x{\rightarrow}\overline{x}_{f}=x\bigg{(}1+\frac{4m_{f}^{2}}{Q^2}\bigg{)}
=\frac{Q^2+4m^2_{f}}{Q^2+W^2} ,
\end{eqnarray}
where $W^2$ denotes the $\gamma^{*}p$ center-of-mass energy
squared. The parameters of the model have been selected from
Ref.[15] as $\sigma_{0}=29.12$, $\lambda=0.277$ and
$x_{0}=0.41{\times}10^{-4}$ for $n_{f}=4$ where the light quark
mass is $m_{f}=0.14~\mathrm{GeV}$ and charm mass is
$1.4~\mathrm{GeV}$.\\
The GBW model\footnote{The GBW model has features of a solution to
the nonlinear evolution of the Balitsky-Kovchegov (BK) type.} was
improved by taking into account the DGLAP evolution of the gluon
density [8]. In the Bartels, Golec-Biernat and Kowalski (BGK)
model [8], the color dipole cross section has the form
\begin{eqnarray}
\sigma_{\mathrm{dip}}(x,\mathbf{r})=\sigma_{0}\bigg{\{}1-
\exp\bigg{(}
\frac{\pi^2r^2\alpha_{s}(\mu^2)xg(x,\mu^2)}{3\sigma_{0}} \bigg{)}
\bigg{\}}.
\end{eqnarray}
The evolution scale $\mu^2$ is connected to the size of the dipole
by $\mu^2=\frac{C}{r^2}+\mu^2_{0}$, and the parameters $C$ and
$\mu_{0}$ are determined from the fits [15] to the HERA data for
$Q^2{\leq}650~\mathrm{GeV}^2$. The gluon density, usually, is
evolved to larger scales $\mu^2$ using the DGLAP evolution
equation at the LO or NLO approximations and its dependent on  the
gluon density parametrized at the starting scale $\mu_{0}^2$ by
the following form
\begin{eqnarray}
\frac{\partial{g(x,\mu^2)}}{\partial{\ln}\mu^2}=P_{gg}(\alpha_{s}(\mu^2),x){\otimes}g(x,\mu^2),
\end{eqnarray}
where $P_{gg}$ is the splitting function and contains real and
virtual terms with the number of active quark flavors $n_{f}$. The
initial conditions of the gluon density are considered in three
forms. The first form is the soft ansatz as used in the original
BGK model and other ones are the soft+hard and soft+negative
ansatzs parametrized in other models [16,17].\\
In another model, the author in Ref.[14] is investigated the
relation between the gluon density obtained using a dipole model
and the standard gluons obtained from the collinear factorization
theorem (CFT). Within the LO $k_{t}$-factorization theory, the
author in Ref.[14] is obtained the longitudinal $\gamma^{*}p$
cross section as
\begin{eqnarray}
\sigma_{L}(x,Q^2){\propto}\int_{0}^{1}dz[z(1-z)]^2\int\frac{d^{2}k_{t}}{k_{t}^4}
\int{d^{2}p}\bigg{(}\frac{1}{\widehat{Q}^{2}+p^2}
-\frac{1}{\widehat{Q}^{2}+(p+k_{t})^2}\bigg{)}^2f(x,k_{t}^{2}),
\end{eqnarray}
where $\widehat{Q}^{2}=z(1-z)Q^2$ and by using the identity
\begin{eqnarray}
\frac{1}{\widehat{Q}^{2}+p^2}=\frac{1}{2\pi}\int{d^2r}\exp(i\mathbf{p}.\mathbf{r})
K_{0}(\widehat{Q}r),
\end{eqnarray}
one can rewrite the unintegrated gluon distribution into the
dipole cross section
\begin{eqnarray}
\sigma(x,\mathbf{r})=\frac{8\pi^{2}}{N_{c}}\int\frac{dk_{t}}{k_{t}^{^{3}}}[1-J_{0}(k_{t}\mathbf{r})]\alpha_{s}f(x,k_{t}^{2}).
\end{eqnarray}
the unintegrated gluon distribution is obtained [14]
\begin{eqnarray}
f(x,k_{t}^{2})=\frac{3\sigma_{0}}{4\pi^2\alpha_{s}}k_{t}^{4}(x/x_{0}^{\lambda})
e^{-k_{t}^{2}(x/x_{0})^\lambda}
\end{eqnarray}
where the integrated gluon distribution is obtained by using
Eq.(3) at fixed coupling in the following form
\begin{eqnarray}
xg(x,Q^{2})=\frac{3\sigma_{0}}{4\pi^2\alpha_{s}}\bigg{[}-Q^2e^{-Q^{2}(x/x_{0})^\lambda}+(x_{0}/x)^\lambda
(1-e^{-Q^{2}(x/x_{0})^\lambda})\bigg{]}.
\end{eqnarray}
Other UGDs relevant at $k_{t}$-factorization have been derived
[1,15,18,19-21], and comparisons between these models have been
analyzed in Refs.[22-23] and summarized in the Appendix.\\

\subsection{III. Method and Results}
$\bullet$ Unintegrated and integrated gluon density:\\

The GBW and BGK models were originally formulated in the
position-space version of the $k_{t}$-factorization formula.
Effects of exact gluon kinematics on the parameters of the GBW and
BGK saturation models were investigated recently in Ref.[24]. The
GBW model, despite implementing gluon saturation, is reasonable at
small transverse momenta $k_{t}$ (after the Fourier transform), as
its exponential decay contradicts the expected perturbative
behavior at large $k_{t}$, while the BGK model is an attempt to
correct that (another known model that corrects the GBW is the
McLerran-Venugopalan model).\\
At present, we consider the GBW and BGK saturation
models\footnote{The behavior of the dipole cross sections in the
GBW and BGK models, at small and large dipoles, are considered in
Ref.[24].} to access the integrated and unintegrated gluon
distribution at low $x$. These models (i.e., GBW and BGK) were
originally formulated in the position-space version of the
$k_{t}$-factorization formula. The GBW and BGK dipole models
preserve very good description of HERA I+II data for every $Q^2$.
We consider the GBW and BGK dipole cross sections and extract the
gluon distribution owing to these models in the following form
\begin{eqnarray}
\alpha_{s}(\mu^2)xg(x,\mu^2)=\frac{3\sigma_{0}}{4\pi^2}Q_{0}^2(\frac{x_{0}}{x})^\lambda
\end{eqnarray}
where the running coupling at the LO approximation is
\begin{eqnarray}
\alpha_{s}(\mu^2)=\bigg{[}
\frac{11C_{A}-2n_{f}}{12\pi}\ln\bigg{(}\frac{\mu^2}{\Lambda_{QCD}^{2}}
\bigg{)}\bigg{]}^{-1},
\end{eqnarray}
and $C_{A}=N_{c}=3$ is the Casimir operator in the fundamental and
adjoint representation of the $\mathrm{SU(N_{c})}$ color group.
Indeed, the gluon density is parametrized at the scale $\mu^2$
using the running coupling $\alpha_{s}(\mu^2)$ as
\begin{eqnarray}
xg(x,\mu^2)=\frac{\sigma_{0}}{16\pi^3}Q_{0}^2(\frac{x_{0}}{x})^\lambda(11C_{A}-2n_{f})\ln\bigg{(}\frac{\mu^2}{\Lambda_{QCD}^{2}}
\bigg{)}.
\end{eqnarray}
The QCD parameter $\Lambda$ is extracted by
$\alpha_{s}(M_{Z}^{2})=0.1166$ using the c-quark threshold.\\
In Fig.1, the results based on the GBW+BGK models for
$\alpha_{s}(\mu^2)xg(x,\mu^2)$ and  $xg(x,\mu^2)$ for $\mu^2=2,
10$ and $100~\mathrm{GeV}^2$ in a wide range of $x$ are plotted.
The results are compared with the results in Ref.[14]. These
results clearly demonstrate that the gluon density extracted from
the GBW and BGK dipole cross sections provides correct behaviors
of the gluon distribution in a wide range of $x$.
\begin{figure}[h]
\centerline{
\includegraphics[width=0.6\textwidth]{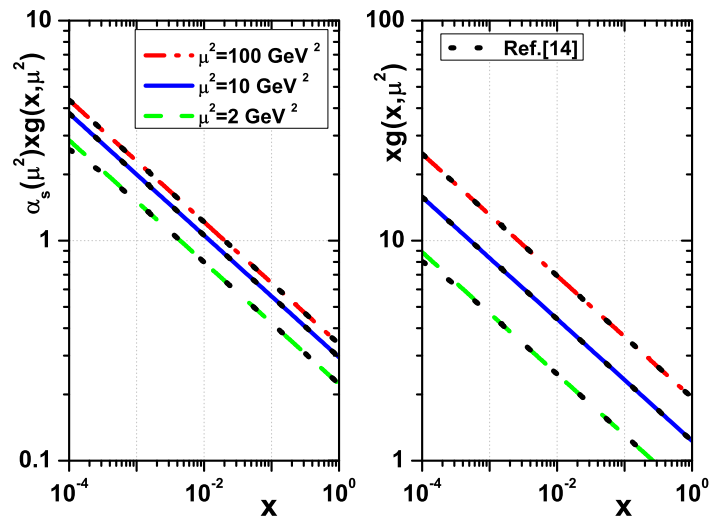}}
\caption{Left: The extracted $\alpha_{s}(\mu^2)xg(x,\mu^2)$ as a
function of $x$ compared with the results in Ref.[14] at $\mu^2=2, 10$ and $100~\mathrm{GeV}^2$. \\
Right: The extracted $xg(x,\mu^2)$ as a function of $x$ compared
with Ref.[14] at $\mu^2=2, 10$ and
$100~\mathrm{GeV}^2$.}\label{Fig1}
\end{figure}
\begin{figure}[h]
\centerline{
\includegraphics[width=0.6\textwidth]{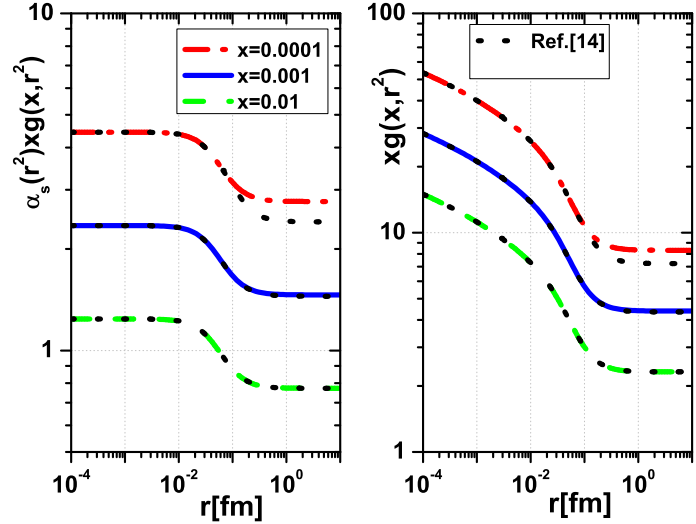}}
\caption{Left: The extracted $\alpha_{s}(\mu^2)xg(x,\mu^2)$ as a
function of $\mathrm{r(fm)}$ compared with the results in Ref.[14] at $x=10^{-2}..10^{-4}$. \\
Right: The extracted $xg(x,\mu^2)$ as a function of
$\mathrm{r(fm)}$ compared with Ref.[14] at
$x=10^{-2}..10^{-4}$.}\label{Fig2}
\end{figure}
In Fig.2, we compared our results with the results in Ref.[14] for
a wide range of the transverse dipole size $r$ at
$x=10^{-2}..10^{-4}$. We observe that the
$\alpha_{s}(\mu^2)xg(x,\mu^2)$ and $xg(x,\mu^2)$ are in a very
good agreement with the results in Ref.[14] in a wide range of $r$
and $x$. We see that there is a deviation between the results at
large $r$ ($r>10^{-1}~\mathrm{fm}$) for $x=10^{-4}$. This behavior
shows
that saturation scale is difference on these models.\\
To realistically describe the structure of the proton, we must
introduce a $k_{t}$ unintegrated gluon density, whose evolution at
low-$x$ is governed by the BFKL equation. The object of the BFKL
evolution equation at very low $x$ is the differential gluon
structure function\footnote{Eq.(22) is modified with the Sudakov
form factor as $x$ increase by the following form, $$
f(x,k_{t}^{2})=\frac{\partial{[xg(x,\mu^{2})S(r,\mu^2)]}}{\partial{\ln}\mu^{2}}|_{\mu^{2}=k^{2}_{t}}.$$
The Sudakov form factor in Ref.[24] is defined by
$S(r,\mu^2)=\frac{\alpha_{s}N_{c}}{4\pi}\ln^{2}(\frac{\mu^2r^2}{4e^{-2\gamma_{E}}})$
where $\gamma_{E}$ is the Euler-Mascheroni constant. } of proton
\begin{eqnarray}
f(x,k_{t}^{2})&=&\frac{\partial{xg(x,\mu^{2})}}{\partial{\ln}\mu^{2}}|_{\mu^{2}=k^{2}_{t}}
\end{eqnarray}
which emerges in the color dipole picture (CDP) of inclusive deep
inelastic scattering (DIS). Unintegrated distributions are
required to describe measurements
where transverse momenta are exposed explicitly.\\
Using the relationship between the dipole cross-section and the
unintegrated gluon distribution (i.e., Eq.(22)) it is
straightforward to obtain
\begin{eqnarray}
f(x,k_{t}^{2})&=&\frac{3\beta_{0}\sigma_{0}}{4\pi^2}Q_{0}^2(\frac{x_{0}}{x})^\lambda\bigg{[}1+\frac{4{\lambda}m_{c}^{2}\ln(\frac{k_{t}^2}{\Lambda_{QCD}^{2}})}
{k_{t}^2(1+\frac{4m_{c}^{2}}{k^{2}_{t}})}
\bigg{]}~{\equiv}~\rho(k^{2}_{t})x^{-\lambda}.
\end{eqnarray}
The resulting UGD with the $k^{2}_{t}$-dependence and comparison
with the HSS [20], IN [18] and WMR [21] models are shown in Figs.
3 and 4. In Figs.3 and 4 we plot the $k^{2}_{t}$ dependence of the
UGD at $x=10^{-3}$ and $10^{-4}$, respectively.
\begin{figure}[h]
\centerline{
\includegraphics[width=0.6\textwidth]{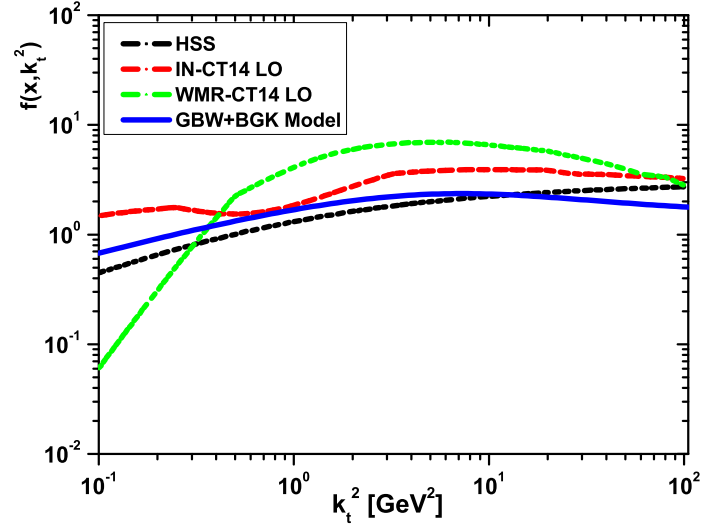}}
\caption{UGD obtained from the GBW+BGK model (solid curve) as a
function of $k_{t}^{2}$  at $x=10^{-3}$, compared with the HSS
[20], IN [18] and WMR [21] models.}\label{Fig3}
\end{figure}
\begin{figure}[h]
\centerline{
\includegraphics[width=0.6\textwidth]{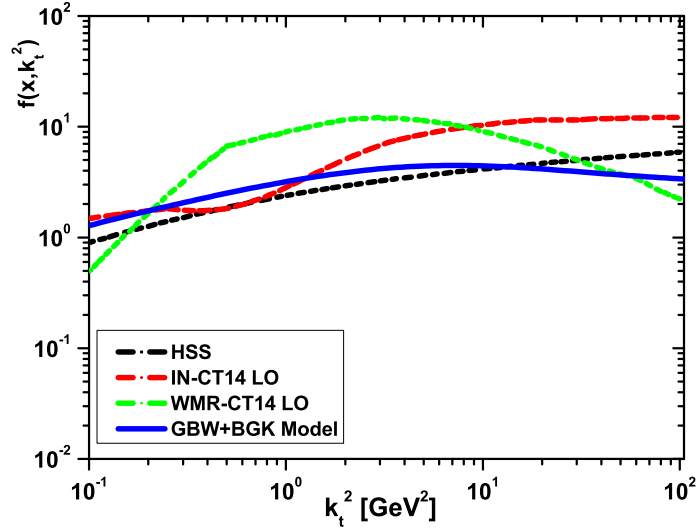}}
\caption{The same as Fig.3 for $x=10^{-4}$.}\label{Fig3}
\end{figure}
In Fig.3, the $k_{t}$ distributions of three different
unintegrated gluons at $x=10^{-3}$  are shown. We observe that the
GBW+BGK saturation model result, in a wide range of $k^{2}_{t}$,
is comparable with the HSS and IN models. There is some
suppression due to the models at large and small values of
$k^{2}_{t}$. The differences are not large in a wide range of
$k^{2}_{t}$ in comparison with the HSS model [20]. In Fig.4 we
compared the GBW+BGK unintegrated gluons with the UGD models for
$x=10^{-4}$ in a wide range of $k_{t}^{2}$. We observe that the
same behavior at low values of
$x$ is consistent with the  UGD models in the $k_{t}$ dependence.\\
\begin{figure}[h]
\centerline{
\includegraphics[width=0.6\textwidth]{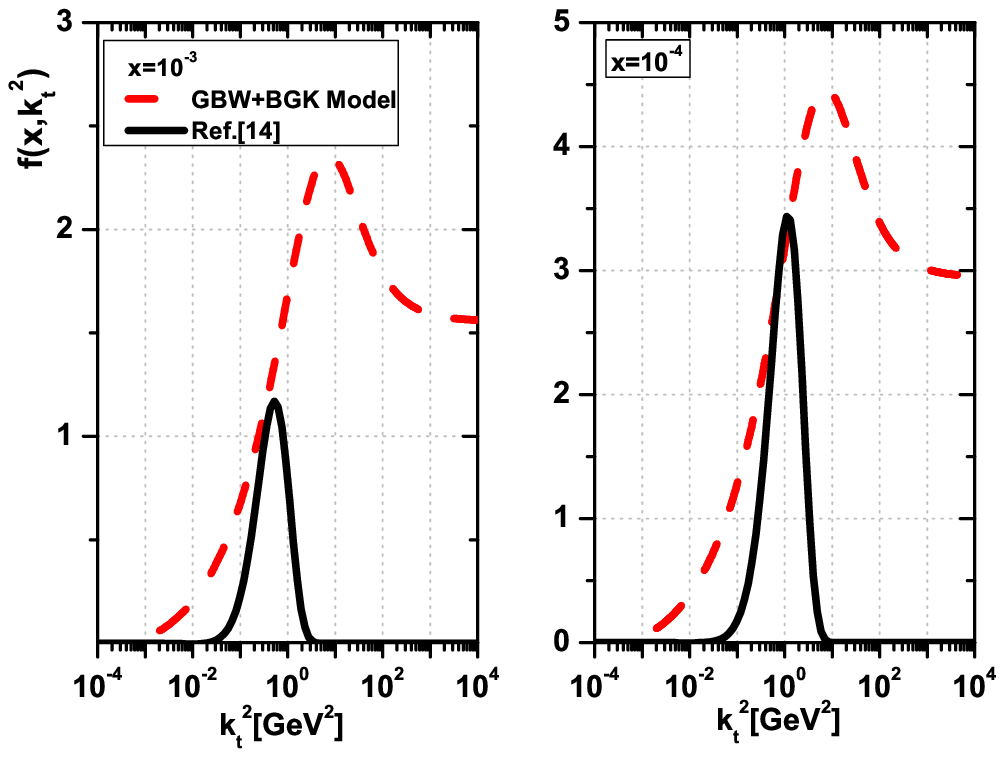}}
\caption{ UGD with the GBW+BGK model as a function of $k_{t}^{2}$
compared with the results in Ref.[14] for $x=10^{-3}$ (left plot)
and $x=10^{-4}$ (right plot).}\label{Fig6}
\end{figure}
In Fig.5, we compared the UGD from the GBW+BGK model with Ref.[14]
in a wide range of $k_{t}^{2}$ for $x=10^{-3}$ and $10^{-4}$
respectively. We observe that $f(x,k_{t}^{2})$ in Ref.[14] is
nearly symmetric. The symmetric point and the maximum of
$f(x,k_{t}^{2})$ increases as $x$ decreases. We can see an
enhancement then a depletion in the UGD of the GBW+BGK model with
increases of $k_{t}$. The maximum of $f(x,k_{t}^{2})$ in the
GBW+BGK model is larger than Ref.[14]. In the limit where $k_{t}$
approaches zero, the behavior of the UGD in the GBW+BGK and the
results in Ref.[14] is similar with a different rate. In the large
$k_{t}$ limit, the UGD behavior in the GBW+BGK model is almost
constant (with uniform rate) where this is similar to the other
UGD models\footnote{In APIPSW model:
$f(x,k_{t}^{2})|_{k_{t}{\rightarrow}\infty}{\simeq}\frac{A}{4\pi^2M^{2}}$\\
In IN model:~~~~~~~~
$f(x,k_{t}^{2})|_{k_{t}{\rightarrow}\infty}{\simeq}F_{hard}.$ }
(such as APIPSW [19] and IN [18] models). Indeed ,at very small
$x$ ($x{\lesssim}10^{-3}$), the scaling violations in the gluon
density are strong. Additionally, for large $k_{t}$, it is
expected that perturbative QCD should accurately describe hard
gluon radiation.\\

$\bullet$ Longitudinal structure function:\\
In the $k_{t}$-factorization, the longitudinal structure function
is driven at low $x$ primarily by gluons and is related in the
following form to the UGD
\begin{eqnarray}
F_{L}(x,Q^2)&=&2\frac{Q^4}{\pi^2}\sum_{f=u,d,s,c}
e_{f}^{2}\int\frac{dk_{t}^2}{k_{t}^4}\int^{1}_{0}d{\beta}\int
d^{2}{\mathbf{\kappa}}'
\alpha_{s}(\mu^2)\beta^2(1-\beta)^2\frac{1}{2}\bigg{(}\frac{1}{D_{1q}}-\frac{1}{D_{1q}}
\bigg{)}^2 f(\frac{x}{z},k_{t}^{2}),
\end{eqnarray}
where $\mathbf{\kappa}'$ is the shifted transverse momentum and
the variable $\beta$ is the Sudakov parameter. They are described
by the sum of the quark box (and crossed box) diagram contribution
to the photon-gluon fusion [25,26] where the photon is
longitudinally polarized. The variables $D_{1q}$ and $D_{2q}$ read
into these parameters as reported in Ref.[26].\\
The on-shell limit of the $k_{t}$-factorization formula (i.e.,
Eq.(24)) can be described by the standard collinear factorization
formula if  the transverse momentum of the gluon is much smaller
than the virtuality of the photon. The expression under the
integral in Eq.(24) can be expanded to the first order in
$k_{t}^{2}$ as
\begin{eqnarray}
\bigg{(}\frac{1}{D_{1q}}-\frac{1}{D_{1q}}
\bigg{)}^2=\frac{4\cos^2{\phi}\kappa'^{2}k_{t}^2}{(\kappa'^{2}+\beta(1-\beta)Q^2+m_{f}^{2})^4}.
\end{eqnarray}
Based on the strong ordering in the transverse momenta, the
on-shell longitudinal structure function at the scale
$\mu^2{\sim}Q^2$ is defined
\begin{eqnarray}
F_{L}(x,Q^2)&=&2\sum_{f=u,d,s,c}{e_{f}^{2}}\bigg{[}
\frac{\alpha_{s}}{\pi}\int_{\overline{x}_{f}}^{1}\frac{dy}{y}(\frac{x}{y})^2(1-\frac{x}{y})
\sqrt{1-\frac{4m_{f}^{2}x}{Q^2(y-x)}}yg(y,Q^2)\nonumber\\
&&-2\frac{m_{f}^{2}}{Q^2}
\frac{\alpha_{s}}{\pi}\int_{\overline{x}_{f}}^{1}\frac{dy}{y}(\frac{x}{y})^3
\ln\bigg{\{}\frac{1+\sqrt{1-\frac{4m_{f}^{2}x}{Q^2(y-x)}}}{1-\sqrt{1-\frac{4m_{f}^{2}x}{Q^2(y-x)}}}
\bigg{\}} yg(y,Q^2) \bigg{]}.
\end{eqnarray}
The authors in Ref.[27] presented a systematic analysis of the
twist expansion in the dipole representation for the inclusive
cross section based on the $k_{t}$-factorization model. the
longitudinal structure function due to the leading twist-2 part in
the dipole picture reads [27,28]
\begin{eqnarray}
F_{L}(x,Q^2)=\frac{1}{3\pi}\sum_{f=u,d,s,c}e_{f}^2\alpha_{s}xg(\xi_{L}x,Q^2),
\end{eqnarray}
where $\xi_{L}{\approx}2$. In order to consider the longitudinal
structure function int the on-shell and the leading twist-2
models, the gluon distribution can be recovered from the GBW+BGK
model (i.e., Eq.21) at the high energy limit. In Fig.6 we show our
results from both models (on-shell and twist-2). We observe that
both are consistent with each other. The description of data, in
the region of low and moderate $Q^2$, is good. The $F_{L}$
structure function is plotted as a function of $x$ in bins of
$Q^2$. Results are compared with H1 data [29,30] and the
parametrization of $F_{L}(x,Q^2)$ [31] at the LO approximation
(CJ15-LO). It is seen that, for all values of the presented $Q^2$,
the extracted longitudinal structure function with respect to the
gluon distribution obtained from the GBW+BGK model, is in
agreement with data. Similar investigations of the longitudinal
structure function have been performed in Refs.[32-37]. In
Ref.[35] an unintegrated gluon density is determined (in a similar
spirit of GBW) and the longitudinal structure function is
calculated and compared to data. In addition, the longitudinal
structure functions in Fig.6 compared with the results
$\mathrm{LLM}^{,}2023$ [35] which is determined according to the
$k_{T}$-factorization and the TMD gluon densities in the
proton\footnote {The longitudinal structure function in Ref.[35]
is defined as a convolution
$$
F_{L}(x,Q^2)=\int_{x}^{1}\frac{dz}{z}\int
d\mathbf{k}^{2}_{T}\sum{e_{f}^{2}}\widehat{C}^{g}_{L}(x/z,Q^2,m_{f}^{2},\mathbf{k}^{2}_{T})
f_{g}(z,\mathbf{k}^{2}_{T},\mu^2),
$$
where the initial TMD gluon distribution is defined in Ref.[38] by
the following form
$$
f_{g}(x,\mathbf{k}^{2}_{T})=c_{g}(1-x)^{b_{g}}\sum_{n=1}^{3}(c_{n}R_{0}(x)|\mathbf{k}_{T}|)^{n}e^{-R_{0}(x)|\mathbf{k}_{T}|}
$$
and the hard coefficient function $\widehat{C}_{L}^{g}$
corresponds to the quark-box diagram for off-shell (dependent on
the incoming gluon virtuality) photon-gluon fusion subprocess in
Ref.[39]}. The LLM [35] results accompanied with the shaded bands
correspond to theoretical uncertainties of the
$k_{T}$-factorization connected with the choice of hard scales.
The predictions are compatible with the $k_{T}$-factorization
within the uncertainties.\\
In Fig.7, we have calculated the $Q^2$-dependence, at low $x$, of
the longitudinal structure function at a fixed value of the
invariant mass $W$, $W=230~\mathrm{GeV}$. Calculations have been
performed from the on-shell model. Since the photon wave function
depends on mass of the quarks in the $q\overline{q}$ dipole, we
considered contributions to the longitudinal structure function
due to the coefficients from the fixed ($n_{f}=4$) and variable
$n_{f}=3+\mathrm{charm}$ active flavors. The fixed
parameters\footnote{The three parameters are $\sigma_{0}$,
$\lambda$ and $x_{0}$.} in Eq.(21) are defined in the first two
rows in Table 1 from Re.[15]. The extracted longitudinal structure
functions are in a good agreement with the H1 data [29],in a wide
range of $Q^2$, in both fixed parameters. We observe that the
results with fixed active flavor number $n_{f}=4$ are smaller than
the results with variable active flavor number
$n_{f}=3+\mathrm{charm}$. However, the both results are comparable
with experimental data in a wide range of
$Q^2$.\\
\begin{figure}[h]
\centerline{
\includegraphics[width=0.75\textwidth]{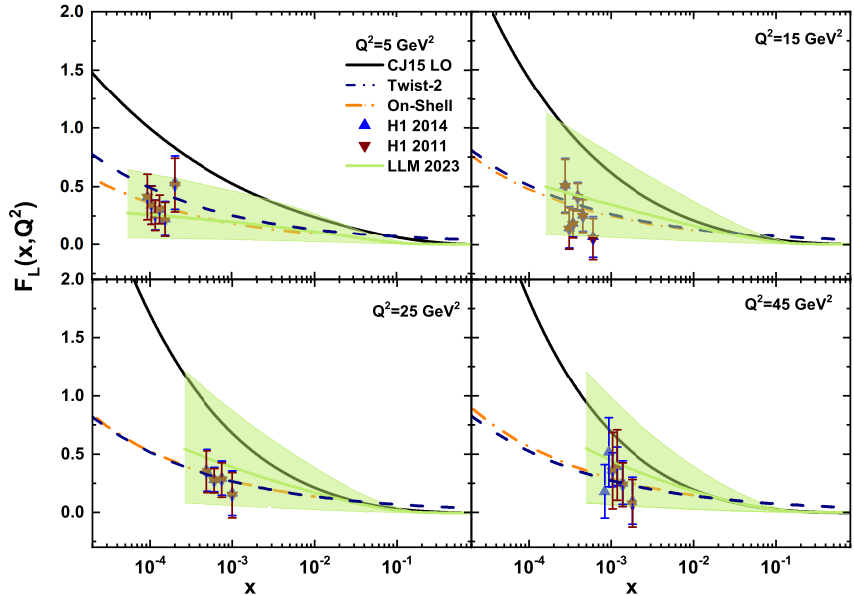}}
\caption{Longitudinal structure function extracted from the
on-shell (dashed-dot curve) and twist-2 (dashed curve) models due
to the gluon distribution obtained form the GBW+BGK model at fixed
$Q^2$ as a function of $x$ variable. The results compared with the
CJ15 [31] parametrization model at the LO approximation and the
experimental data (up-triangle H1 2014 [29], down-triangle H1 2011
[30]) as accompanied with total errors. In addition, the results
compared with the $\mathrm{LLM}^{,}2023$ [35]
$k_{T}$-factorization with scale uncertainties (the green
bands).}\label{Fig7}
\end{figure}
\begin{figure}[h]
\centerline{
\includegraphics[width=0.6\textwidth]{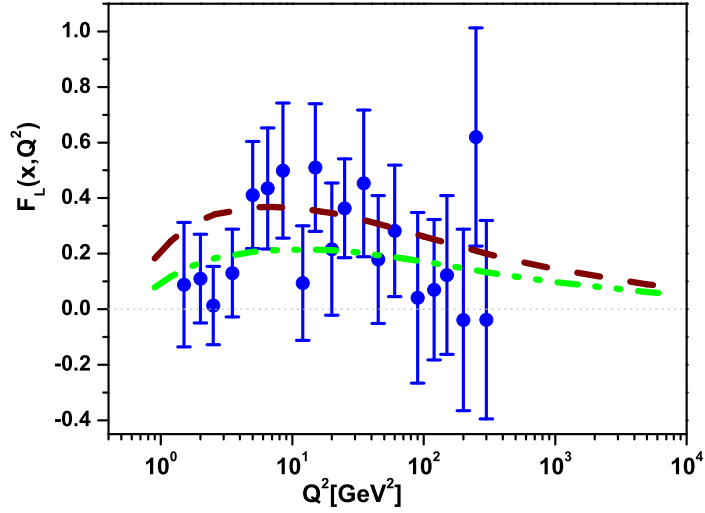}}
\caption{Longitudinal structure function extracted from the
on-shell model due to the gluon distribution obtained form the
GBW+BGK model as a function of variable $Q^2$ at fixed value of
the invariant mass  $\mathrm{W}=230~\mathrm{GeV}$. The results
calculated with the fixed parameters in the dipole model [15] for
$n_{f}=4$ (dashed-dot curve) and $n_{f}=3+\mathrm{charm}$ (dashed
curve). Experimental data are from the H1-Collaboration [29] as
accompanied with total errors.}\label{Fig7}
\end{figure}

\subsection{IV. Conclusion}

In this paper we derive results for unintegrated color dipole
gluon distribution function at small transverse momentum in the
GBW+BGK model. We have shown that the $k_{t}$ factorized,
evaluated within the GBW+BGK model, describes the longitudinal
structure functions in the small-$x$ region, very well. In the
first part of this analysis, we obtained a new description for the
unintegrated gluon distribution which is close to the results in
Ref.[14] for a wide range of the transverse dipole size $r$ at low
$x$. We have shown, the behavior of the GBW+BGK model for the UGD
is comparable with the UGD models in a wide range of $k_{t}^{2}$.
In the second part of our study, we have obtained a good agreement
of the proton structure function $F_{L}$, using the GBW+BGK model
within the $k_{t}$-factorization framework for both on-shell and
twist-2 corrections. We have analyzed the impact of the exact
kinematics in the $k_{t}$-factorization scheme as they are very
important for the phenomenological description of the data on
$F_{L}$. In particular, it leads to larger differences in the
longitudinal structure function when we consider a good fit
quality with the charm and light quark contributions to the dipole
coefficients. Therefore, we conclude that the GBW+BGK model
provides an economical description of the data on the longitudinal
structure function  for fixed $W$. Explicit, analytical
expressions for the integrated and unintegrated gluon distribution
in the GBW+BGK model are obtained in terms of the effective
parameters of the CDP and results of numerical calculations of the
$F_{L}$ as well as comparisons with available experimental are
presented.\\

\subsection{ACKNOWLEDGMENTS}
The author is grateful to Razi University for the financial
 support of this project. G.R.Boroun thanks A.V. Lipatov for
allowing access to data related to the longitudinal structure function with the TMD gluon density.\\

\subsection{APPENDIX}

$\bullet$~$\mathbf{\mathrm{ABIPSW}~ \mathrm{model}:}$\\
This is a $x$-independent model of the UGD which has been proposed
by the authors in Ref.[19] as
\begin{eqnarray}
f(x,k_{t}^{2})=\frac{A}{4\pi^{2}M^{2}}\Big{[}\frac{k_{t}^{2}}{M^{2}+k_{t}^{2}}\Big{]},
\end{eqnarray}
where merely coincides with the proton impact factor. Here M is a
characteristic soft scale and A is the normalisation factor.\\

$\bullet$~$\mathbf{\mathrm{IN}~ \mathrm{model}:}$\\
In the large and small $k_{t}$ regions, a UGD soft-hard model
(where the soft and the hard components are defined in [18]) has
been proposed by the authors in Ref. [18] as
\begin{eqnarray}
f(x,k_{t}^{2})=f_{\mathrm{soft}}^{(B)}(x,k_{t}^{2})\frac{k_{s}^{2}}{k_{s}^{2}+k_{t}^{2}}
+f_{\mathrm{\mathrm{hard}}}(x,k_{t}^{2})\frac{k_{t}^{2}}{k_{h}^{2}+k_{t}^{2}}.
\end{eqnarray}

$\bullet$~$\mathbf{\mathrm{HSS}~ \mathrm{model}:}$\\
This model [20] is used in the study of DIS structure functions
and takes the form of a convolution between the BFKL gluon
Green$^{,}$s function and a leading-order (LO) proton impact
factor, where has been employed in the description of
single-bottom quark production at LHC and to investigate the
photoproduction of $J/\Psi$ and $\Upsilon$, by the following form
\begin{eqnarray}
f(x,k_{t}^{2},M_{{h}})&=&\int_{-\infty}^{+\infty}\frac{d\nu}{2\pi^{2}}\mathcal{C}
\frac{\Gamma(\delta-i\nu-\frac{1}{2})}{\Gamma(\delta)}(\frac{1}{x})^{\chi(\frac{1}{2}+i\nu)}
(\frac{k_{t}^{2}}{Q_{0}^{2}})^{\frac{1}{2}+i\nu}\Bigg{\{}1+\frac{\overline{\alpha}_{s}^{2}\beta_{0}\chi_{0}(\frac{1}{2}+i\nu)}{8N_{c}}\log(\frac{1}{x})\nonumber\\
&&{\times}\Bigg{[}-\psi(\frac{1}{2}+i\nu)-\log\frac{k_{t}^{2}}{M^{2}_{{h}}}\Bigg{\}}\Bigg{]}.
\end{eqnarray}
In the above equation (i.e., Eq.(30)),
$\chi_{0}(\frac{1}{2}+i\nu)$ and $\chi(\gamma)$ are respectively
the LO and the next-to-leading order (NLO) eigenvalues of the BFKL
kernel and $\beta_{0}=11-\frac{2}{3}n_{f}$ with $n_{f}$ the number
of active quarks. Here
$\overline{\alpha}_{s}=\frac{3}{\pi}\alpha_{s}(\mu^{2})$ with
$\mu^{2}=Q_{0}M_{h}$ where $M_{h}$ plays the role of the hard
scale which can be identified with the photon virtuality,
$\sqrt{Q^{2}}$.\\

$\bullet$~$\mathbf{\mathrm{WMR}~ \mathrm{model}:}$\\
 The WMR model [21] depends on an extra-scale $\mu$, fixed at $Q$, by the following
form
\begin{eqnarray}
f(x,k_{t}^{2},\mu^{2})&=&T_{g}(k_{t}^{2},\mu^{2})\frac{\alpha_{s}(k_{t}^{2})}{2\pi}
\int_{x}^{1}dz\Bigg{[}\sum_{q}P_{gq}(z)\frac{x}{z}q(\frac{x}{z},k_{t}^{2})
+P_{gg}(z)\frac{x}{z}g(\frac{x}{z},k_{t}^{2})\Theta(\frac{\mu}{\mu+k_{t}}-z)
\Bigg{]},
\end{eqnarray}
where $T_{g}(k_{t}^{2},\mu^{2})$ gives the probability of evolving
from the scale $k_{t}$ to the scale $\mu$ without parton emission
and $P_{ij}^{,}$s are the splitting functions.\\

$\bullet$~$\mathbf{\mathrm{GBW}~ \mathrm{model}:}$\\
This model [1] derives from the effective dipole cross section
$\sigma(x,\mathbf{r})$ for the scattering of a $q\overline{q}$
pair of a nucleon by the following form
\begin{eqnarray}
f(x,k_{t}^{2})&=&k_{t}^{4}\sigma_{0}\frac{R_{0}^{2}(x)}{2\pi}e^{-k_{t}^{2}R_{0}^{2}(x)},
\end{eqnarray}
with
$R_{0}^{2}(x)=\frac{1}{\mathrm{GeV}^{2}}(\frac{x}{x_{0}})^{\lambda}$.\\


\section{References}
1. K.Golec-Biernat and  M.W$\ddot{\mathrm{u}}$sthoff, Phys. Rev. D
{\bf59},  014017 (1998); K.Golec-Biernat, Acta.Phys.Polon.B
{\bf33}, 2771 (2002);
Acta.Phys.Polon.B {\bf35}, 3103 (2004); J.Phys.G {\bf28}, 1057 (2002).\\
2. J.R.Forshaw and G.Shaw, JHEP {\bf12}, 052 (2004).\\
3. TMD Handbook, R.Boussarie et al, arXiv [hep-ph]: 2304.03302;
A.V.Lipatov, G.I.Lykasov and M.A. Malyshev, Phys.Lett.B {\bf839}, 137780 (2023).\\
4. Yu.L.Dokshitzer, Sov.Phys.JETP{\bf46}, 641 (1977); G.Altarelli
and G.Parisi, Nucl.Phys.B {\bf126}, 298 (1977); V.N.Gribov and
L.N.Lipatov, Sov.J.Nucl.Phys. {\bf15},
438 (1972).\\
5. V.S.Fadin, E.A.Kuraev and L.N.Lipatov, Phys.Lett.B \textbf{60},
50(1975); L.N.Lipatov, Sov.J.Nucl.Phys. \textbf{23}, 338(1976);
I.I.Balitsky and L.N.Lipatov, Sov.J.Nucl.Phys.
\textbf{28}, 822(1978).\\
6. R.Boussarie and Y.Mehtar-Tani, Phys.Lett.B {\bf831}, 137125 (2022).\\
7. E.Iancu,K.Itakura and S.Munier, Phys.Lett.B {\bf590}, 199
(2004).\\
8. J.Bartels, K.Golec-Biernat and H.Kowalski, Phys. Rev. D{\bf66},
014001 (2002).\\
9. B.Sambasivam, T.Toll and T.Ullrich, Phys.Lett.B {\bf803}, 135277 (2020).\\
10. J.R.Forshaw and G.Shaw, JHEP {\bf12},
052 (2004).\\
11. E.Iancu, A.Leonidov and L.McLerran, Nucl.Phys.A {\bf692}, 583
(2001); Phys.Lett.B {\bf510}, 133 (2001).\\
12. An Assessment of U.S. Based Electron-Ion Collider Science. The
National Academies Press, Washington, DC, 2018.\\
13. P.Agostini et al. [LHeC Collaboration and FCC-he Study Group
], J. Phys. G: Nucl. Part. Phys. {\bf48}, 110501 (2021).\\
14. R.S. Thorne, Phys.Rev.D{71}, 054024 (2005).\\
15. K. Golec-Biernat and S.Sapeta, JHEP
{\bf03}, 102 (2018).\\
16. A.Luszczak and H.Kowalski, Phys.Rev.D {\bf89}, 074051 (2014).\\
17. A.Luszczak, M.Luszczak and W.Schafer, Phys.Lett.B {\bf835}, 137582 (2022).\\
18. I.P.Ivanov and N.N.Nikolaev, Phys.Rev.D {\bf65}, 054004
(2002).\\
19. I.V. Anikin, A. Besse, D.Yu. Ivanov, B. Pire, L. Szymanowski
and S. Wallon, Phys. Rev. D {\bf84}, 054004 (2011).\\
20. M. Hentschinski, A. Sabio Vera and C. Salas, Phys. Rev. Lett.
{\bf110},  041601 (2013).\\
21. G. Watt, A.D. Martin and M.G. Ryskin, Eur. Phys. J. C {\bf31},
73
(2003).\\
22. A.D.Bolognino, F.G.Celiberto, Dmitry Yu. Ivanov and A.Papa,
arXiv [hep-ph]:1808.02958; arXiv [hep-ph]:1902.04520; arXiv
[hep-ph]:1808.02395; F.G.Celiberto, Nuovo Cim. C {\bf42}, 220
(2019); F.G.Celiberto, D. Gordo Gomez and A.Sabio Vera,
Phys.Lett.B
{\bf786}, 201 (2018).\\
23. G.R.Boroun, Eur.Phys.J.C {\bf83}, 42 (2023);  Eur.Phys.J.C
{\bf82}, 740 (2022); Phys.Rev.D {\bf108}, 034025 (2023).\\
24. T.Goda,K.Kutak and S.Sapeta, arXiv[hep-ph]:2305.14025.\\
25. H. Jung, A.V.Kotikov, A.V.Lipatov and N.P.Zotov,
    Proc. of 15th Int. Workshop on Deep-Inelastic Scattering
    and Related Subjects, Munich, April 2007,
arXiv[hep-ph]:0706.3793v2.\\
26. K.Golec-Biernat and A.M.Stasto, Phys.Rev.D {\bf80}, 014006
(2009).\\
27. J. Bartels, K. J. Golec-Biernat and K. Peters, Eur. Phys. J. C
{\bf17}, 121 (2000).\\
28. N.N.Nikolaev and B.G.Zakharov, Phys.Lett.B {\bf327}, 149 (1994); Phys.Lett.B {\bf332}, 184 (1994).\\
29. V.Andreev, A.Baghdasaryan, S.Baghdasaryan, et al.(H1
Collab.), Eur.Phys.J.C {\bf74}, 2814 (2014).\\
30. F.D.Aaron, C.Alexa, V.Andreev, et al.(H1 Collab.), Eur.Phys.J.C {\bf71}, 1579 (2011).\\
31. A.Accardi, L.T.Brady, W.Melnitchouk, J.F.Owens and N.Sato,
Phys.Rev.D {\bf93}, 114017 (2016).\\
32. L.P.Kaptari, A.V.Kotikov, N.Yu.Chernikova and P.Zhang,
Phys.Rev.D {\bf99}, 096019 (2019).\\
33. S.Zarrin and S.Dadfar, Int.J.Theor.Phys. {\bf60}, 3822
(2021).\\
34. G.R.Boroun, Phys.Rev.D {\bf105}, 034002 (2022).\\
35. A.V.Lipatov, G.I.Lykasov and M.A.Malyshev, Phys.Lett.B
{\bf839}, 137780 (2023).\\
36. R.Saikia, P.Phukan and J.K.Sarma, arXiv [hep-ph]:2304.00272.\\
37. Z.B.Baghsiyahi, M.Modarres and R.K.Valeshabadi, Eur.Phys.J.C
{\bf82}, 392 (2022).\\
38. A.V.Lipatov, G.I.Lykasov, M.A.Malyshev, Phys.Rev.D {\bf107},
014022 (2023).\\
39. A.V.Kotikov, A.V.Lipatov, G.Parente and N.P.Zotov,
Eur.Phys.J.C {\bf26}, 51 (2002).\\

\end{document}